# Uncovering coordinated cross-platform information operations: Threatening the integrity of the 2024 U.S. presidential election

by Marco Minici, Federico Cinus,
Luca Luceri, and Emilio Ferrara

## Abstract

Information operations (IOs) pose a significant threat to the integrity of democratic processes, with the potential to influence election-related online discourse. In anticipation of the 2024 U.S. presidential election, we present a study aimed at uncovering the digital traces of coordinated IOs on X (formerly Twitter). Using our machine learning framework for detecting online coordination, we analyze a dataset comprising election-related conversations on X from May to July 2024. This reveals a network of coordinated inauthentic actors, displaying notable similarities in their link-sharing behaviors. Our analysis shows concerted efforts by these accounts to disseminate misleading, redundant, and biased information across the Web through a coordinated cross-platform information operation: The links shared by this network frequently direct users to other social media platforms or mock news sites featuring low-quality political content and, in turn, promoting the same X and YouTube accounts. Members of this network also shared deceptive images generated by AI, accompanied by language attacking political figures and symbolic imagery intended to convey power and dominance. While X has suspended or restricted a subset of these accounts, 75 percent of the coordinated network remains active, garnering substantial traction over time: The suspicious Web sites promoted by this coordinated network are shared thousands of times per day by the X user base, further amplifying their reach and potential impact. Our findings underscore the critical role of developing computational models to scale up the detection of threats on large social media platforms, and emphasize the broader implications of these techniques to detect IOs across the wider Web.

**Contents**

## Introduction

Over the past decade, malicious actors have increasingly sought to manipulate public discourse through information operations (IOs) on social media platforms. The actors behind these IOs employ various techniques, including deploying inauthentic accounts like bots (Ferrara. 2020; Luceri, Deb, Giordano. *et al.*, 2019; Pozzana and Ferrara, 2020), state-sponsored human agents (Badawy, *et al.*, 2019; Bail, *et al.*, 2020), or a combination of both in coordinated networks (Hristakieva, *et al.*, 2022; Nizzoli, *et al.*, 2021; Starbird, *et al.*, 2019), to disseminate misleading content and propaganda at scale. Furthermore, fringe platforms have facilitated the circulation of questionable content, often amplified through duplicate messages, links, and images (Zelenkauskaite, *et al.*, 2021). To aggravate the problem, recent advances in generative AI (GenAI) and its democratization have raised concerns about potential malicious misuse (Luceri, *et al.*, 2024; Mozes, *et al.*, 2023), such as the creation of synthetic personas and credible fake content. GenAI has the potential to dramatically enhance both the effectiveness and credibility of IOs (Ferrara. 2024a, 2024b).

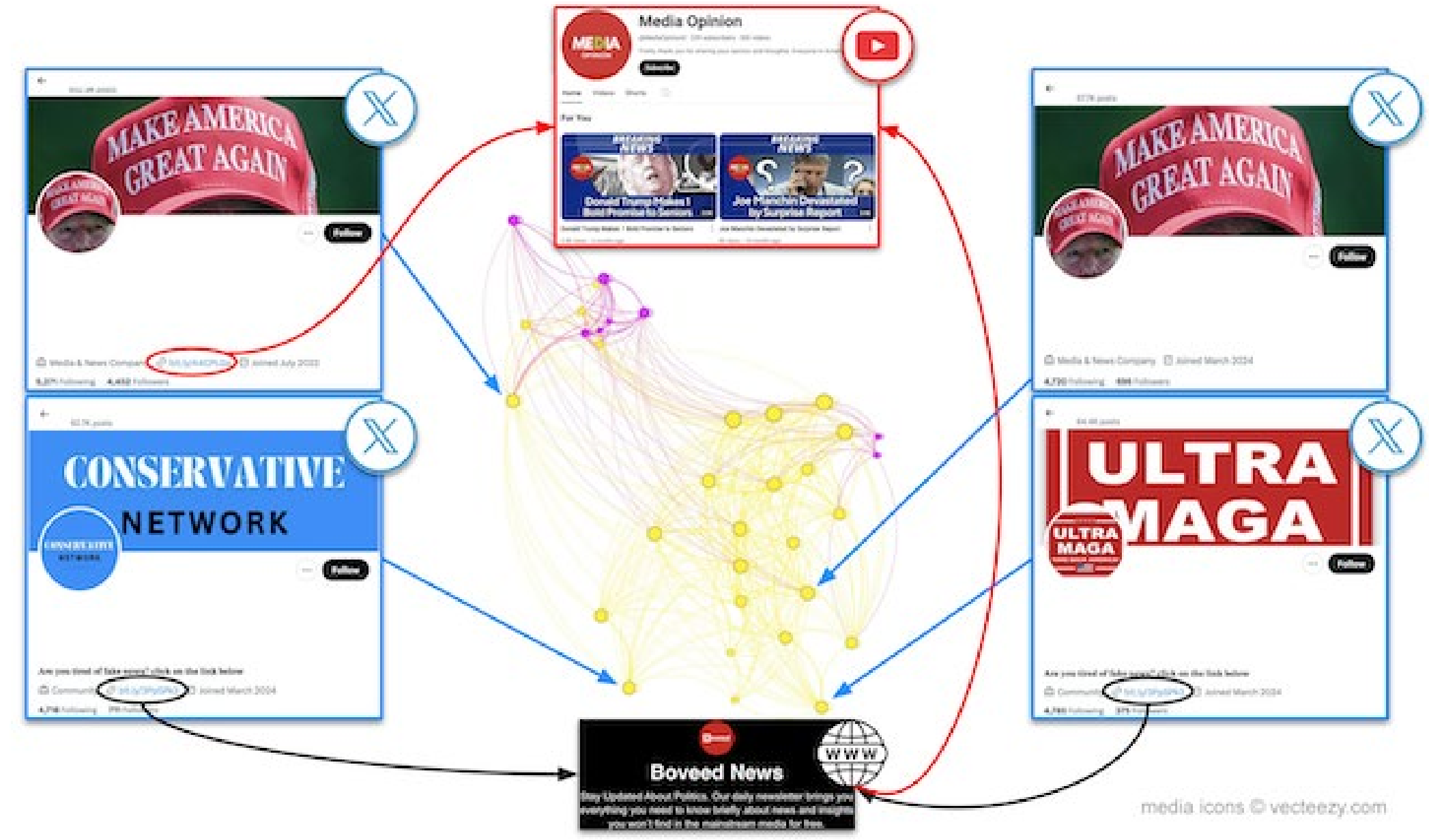

**Figure 1:** Coordinated cross-platform information operation (IO) network across X, YouTube, and the Web. Coordinated accounts on X share links to various news providers and the @MediaOpinion0 YouTube channel. The news providers also link back to the YouTube channel and various X accounts.

Note: Larger version of Figure 1 available here.

IO activities are often strategically timed to coincide with major political events, where public attention is heightened, and the potential impact is greatest (Martin, *et al.*, 2019). For example, during the 2016 U.S. presidential election, the Russian Internet Research Agency (IRA) orchestrated a large-scale disinformation campaign aimed at manipulating public opinion and fostering distrust in the electoral process (Badawy, *et al.*, 2019; Bail, *et al.*, 2020; Bovet and Makse, 2019). Similar operations were detected during the 2018 U.S. midterm elections (Deb,, *et al.*, 2019; Luceri, Deb, Badawy, *et al.*, 2019), the 2020 U.S. presidential election (Ferrara, *et al.*, 2020; Luceri, *et al.*, 2021), and leading up to the 6 January 2021, attack on the U.S. Capitol (Vishnuprasad, *et al.*, 2024). IOs have also been linked to global political events, such as the Ukraine-Russia war (Chen and Ferrara. 2023; Pierri, *et al.*, 2023; Soares, *et al.*, 2023), the conflict in Gaza (Dey, *et al.*, 2024), and protests in Spain, Turkey, Serbia, and Venezuela (Guo and Vosoughi, 2022; Stella, *et al.*, 2018).

(a) Hashtags Statistics

| Hashtag | Count |
|---|---|
| #GOP | 423,427 |
| #JoeBiden | 358,955 |
| #TheDemocrats | 208,307 |
| #POTUS | 199,869 |
| #elonmusk | 146,573 |
| #realDonaldTrump | 139,868 |
| #YouTube | 127,044 |
| #harryjsisson | 108,825 |
| #KamalaHarris | 103,781 |
| #GuntherEagleman | 101,248 |

(b) Mention Statistics

| Mention | Count |
|---|---|
| @MAGA | 430,493 |
| @Trump2024 | 246,652 |
| @BidenHarris2024 | 150,954 |
| @Biden | 123,121 |
| @Trump | 121,629 |
| @maga | 58,223 |
| @Biden2024 | 34,134 |
| @DonaldTrump | 31,908 |
| @JoeBiden | 24,061 |
| @GOP | 23,615 |

(c) Domain Statistics

| Domain | Count |
|---|---|
| foxnews.com | 43,964 |
| l.smartnews.com | 27,091 |
| breitbart.com | 26,443 |
| newsbreak.com | 25,239 |
| rumble.com | 19,719 |
| nytimes.com | 18,234 |
| apple.news | 17,058 |
| nypost.com | 15,212 |
| rawstory.com | 14,764 |
| cnn.com | 13,241 |

**Table 1:** Summary statistics of the full dataset. The top 10 hashtags and mentions appear balanced across the political spectrum. The top 10 web domains consist of widely recognized popular Web sites.

The need to detect IOs on social media has thus become a critical research priority for preserving the integrity of online discourse and democratic processes. Researchers are actively developing advanced detection methods, with a focus on identifying temporal patterns in activity (Magelinski, *et al.*. 2022; Sharma, *et al.*, 2022; Tardelli, *et al.*, 2024; Vargas, *et al.*, 2020; Vishnuprasad, *et al.*, 2024) and co-activity behavior (Luceri, *et al.*, 2024; Nizzoli, *et al.*, 2021; Pacheco, *et al.*, 2020; Pacheco, *et al.*, 2021; Sharma, *et al.*, 2021; Weber and Neumann, 2021). In parallel, private organizations are addressing IO threats on their platforms: OpenAI, for instance, recently uncovered state-sponsored IO campaigns from Russia, China, and Iran, leveraging GenAI to propagate disinformation (Augenstein, *et al.*, 2024; OpenAI, 2024). Similarly, Facebook's transparency report detailed the removal of six IO campaigns tied to Russia, Vietnam, and the U.S. (Meta, 2024), while Microsoft identified an Iranian IO that emerged following the start of the Israel-Hamas conflict (Microsoft Threat Intelligence, 2024). The U.S. government has also acknowledged the growing threat, revealing how Russian state media recruited influencers to spread content on social platforms (U.S. Department of State. Office of the Spokesperson, 2024).

### *Contributions of this work*

In light of the urgency to protect the integrity of online debate in the context of the 2024 U.S. election, we leveraged our scalable IO detection approach (Luceri, *et al.*, 2024) to analyze election-related conversations on X, aiming to uncover potential IOs targeting election discourse. We make the following contributions:

- We deployed our scalable IO detection tool on a large sample of X activity surrounding the 2024 U.S. presidential election, identifying a suspicious network of coordinated actors sharing links to external Web sites.
- We characterized this network's actors and their sharing activity, revealing an orchestrated cross-platform effort to artificially amplify right-leaning narratives through repetitive messaging, generative AI images, and the use of mock, duplicated, or suspicious Web sites.
- We analyzed the temporal evolution of this network of coordinated actors and their sharing activity from May to July 2024, uncovering a growing volume of tweets pointing to suspicious Web domains and increased engagement from the broader user base with these promoted Web sites.

## Data collection

To analyze the online discourse surrounding the 2024 U.S. election, we continuously collected a large dataset of public tweets accessible via X's Web interface [1]. The data was gathered by tracking, in real-time, specific election-related keywords (see the Appendix). This methodology allowed us to capture a wide range of user-generated content related to the election. The data collection spanned from 1 May to 31 July 2024. Our dataset, whose top hashtags, mentions, and Web domains are summarized in Table 1, is representative of the political and election-related conversation during this time span. It comprises 17,826,799 distinct tweets shared by 2,699,777 unique users.

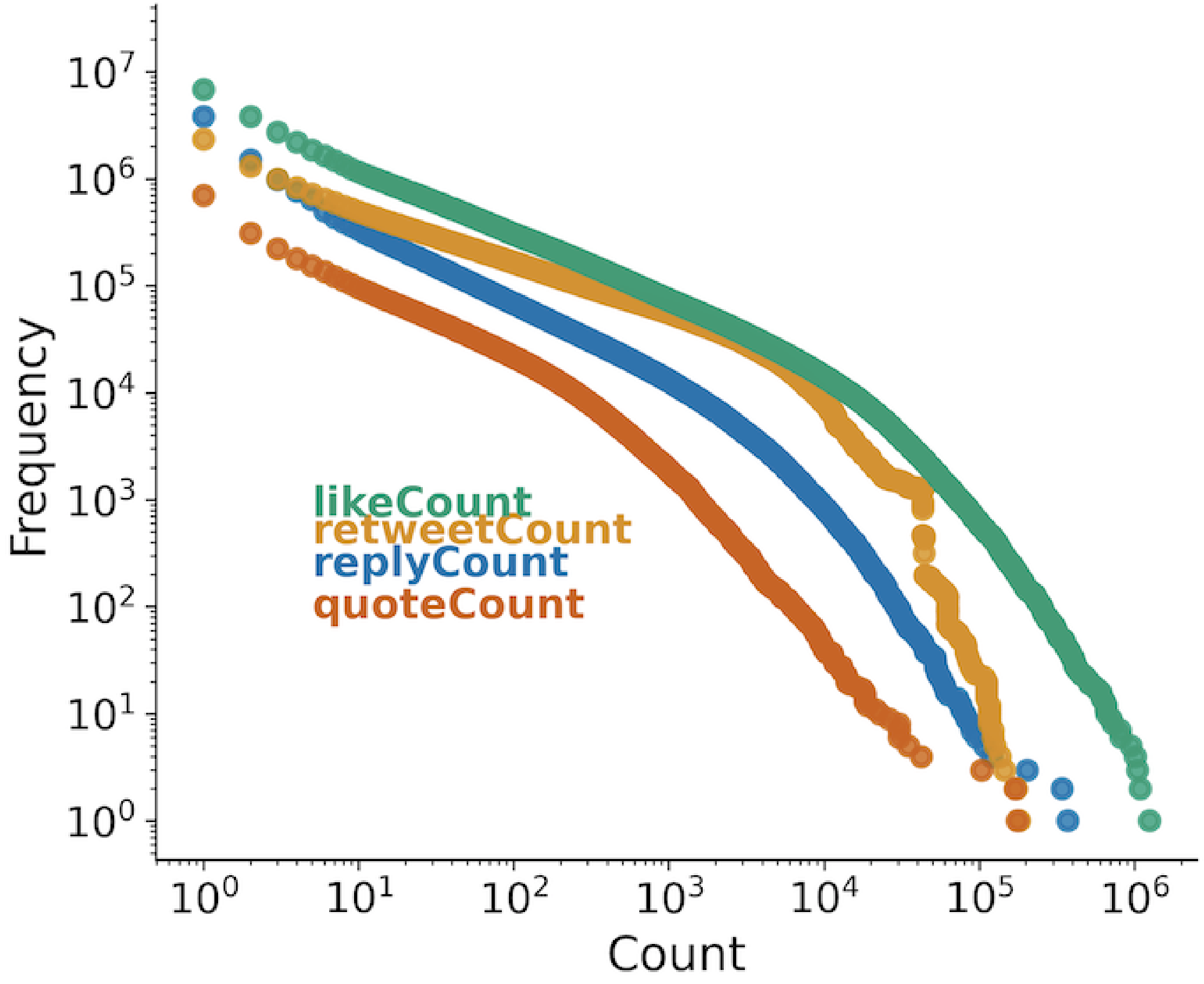

**Figure 2:** Distribution of user interactions.

As shown in Tables 1(a) and (b), the most frequent hashtags and mentions pertain to the above mentioned political topics. Hashtags like #GOP and #JoeBiden dominate the conversation, while mentions of political accounts such as @MAGA and @Trump2024 are also prominent. This collection of tweets provides an essential resource for analyzing user behavior, engagement with political content, and patterns of interaction on social media throughout the period under scrutiny.

We report in Figure 2 the distributions of the following forms of interactions: likes, retweets, replies, and quotes. Interactions resemble a power-law distribution which is expected and corroborates the data collection.

## Methodology

Building on an established approach in the literature, we leverage co-activity patterns to extract groups of coordinated users (Luceri, *et al.*, 2024; Magelinski, *et al.*, 2022; Nizzoli, *et al.*, 2021; Pacheco, *et al.*, 2020; Pacheco, *et al.*, 2021; Tardelli, *et al.*, 2024; Vargas, *et al.*, 2020; Vishnuprasad, *et al.*, 2024; Weber and Neumann, 2021). In particular, in this report, we focus on the co-sharing of the same links (or URLs), since prior work unveiled that this pattern is particularly informative of coordination efforts (Gabriel, *et al.*, 2023; Giglietto, *et al.*, 2020; Luceri, *et al.*, 2024). Our method (Luceri, *et al.*, 2024) constructs a bipartite graph that connects users to URLs based on their sharing activities, with weights assigned using TF-IDF to account for the popularity of each URL. Consequently, each user is represented as a TF-IDF vector of the URLs that they shared. This bipartite graph is then transformed into a similarity network, where users are connected based on the similarity of their shared URLs, with connection weights determined by the cosine similarity between the TF-IDF vectors. From this point forward, we refer to the resulting graph as the *Similarity Network*.

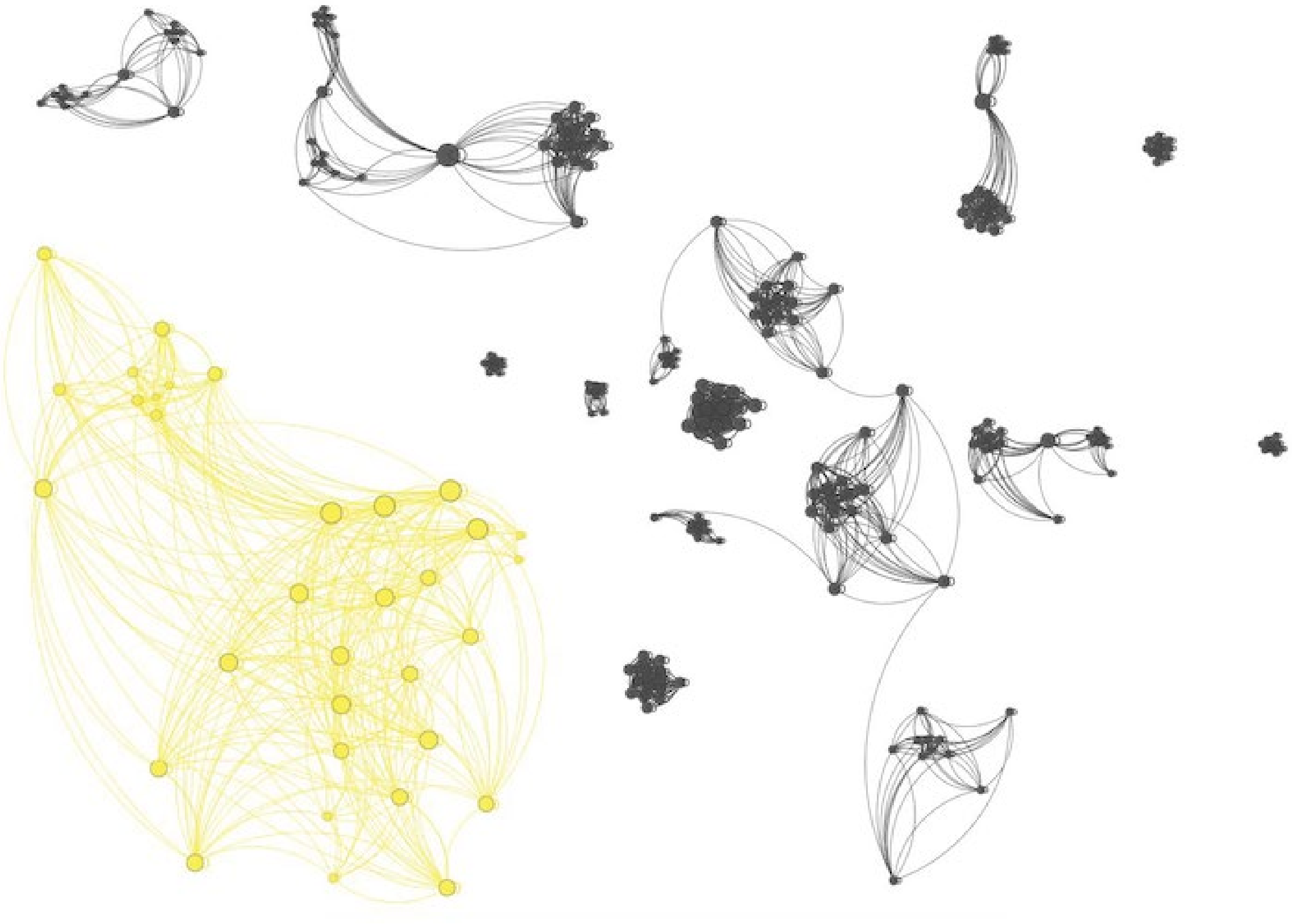

**Figure 3:** CoURL 10-core Similarity network built using our method (Luceri, *et al.*, 2024). The coordinated accounts that we detected are highlighted in yellow.

Note: Larger version of Figure 3 available here.

Our method utilizes eigenvector centrality of users within this similarity network to filter out organic users and identify coordinated inauthentic accounts. The core idea is that IOs, which involve multiple accounts, tend to exhibit a significant degree of collective similarity. In a similarity network, this is illustrated by a node (representing an IO actor) connecting to numerous other nodes. We tested this approach on several verified IOs occurred in the recent past (Luceri, *et al.*, 2024), and found results demonstrating very high precision (>99 percent) when a conservative centrality threshold was applied.

## Results

### *Identifying coordinated networks of inauthentic actors*

We begin our analysis by extracting the *Similarity Network* from users engaged in the U.S. election debate in May 2024, using the previously described methodology. This network comprises approximately ~ 2*K* nodes and ~ 7*K* edges. Following Luceri, *et al.* (2024), we retained only nodes with an eigenvector centrality greater than the 99th percentile of the centrality distribution to pinpoint networks of (intentionally) coordinated actors. This conservative approach, consistent with the strategies adopted in previous work to minimize false positives (Luceri, *et al.*, 2024; Pacheco, *et al.*, 2021), allows us to identify a network of 34 highly coordinated accounts. Figure 3 shows the 10-core decomposition of the *Similarity Network*, and Figure 1 shows a zoom-in on the detected cluster of highly coordinated accounts. In particular, only eight out of the 34 coordinated accounts have been suspended by X at the time of writing (25 October 2024), and two are restricted due to unusual activity (see the Appendix). In Figure 3, we highlight in yellow the community of accounts identified as coordinated. In contrast, nodes in black represent non-coordinated users. The absence of connections between components signifies no similarity among them, as each edge denotes a similarity between two users. Coordinated accounts (yellow) form a distinct, densely connected component, clearly separated from other groups. This indicates that coordinated accounts tend to exhibit unusually high similarities in their sharing patterns and form a dense network structure. These signals suggest inauthentic coordination, as supported by previous literature (Gabriel, *et al.*, 2023; Giglietto, *et al.*, 2020; Pacheco, *et al.*, 2021) and our prior research on multiple verified Twitter IOs (Luceri, *et al.*, 2024).

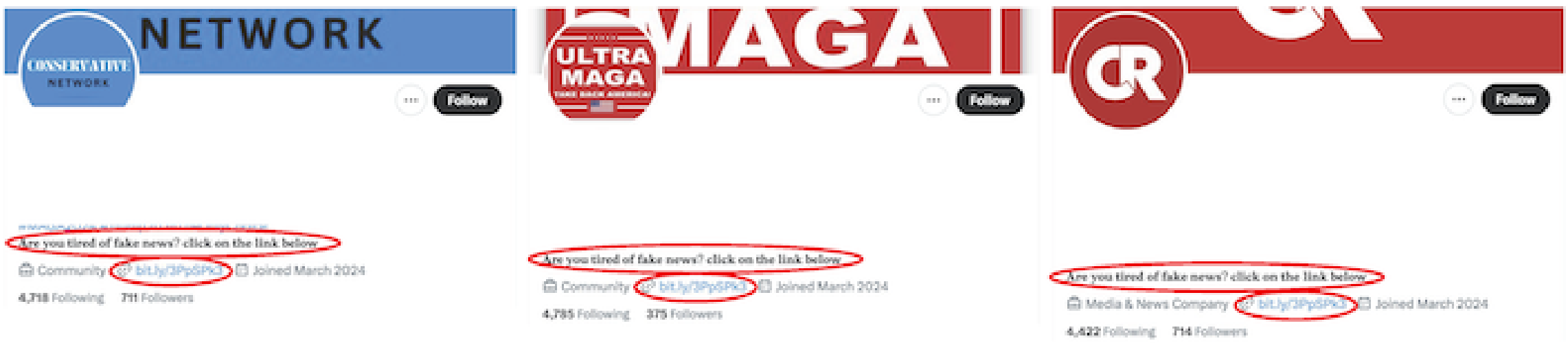

**Figure 4:** Example of three suspicious X accounts with similar descriptions, account aesthetics, and links to the same external domain. Account usernames have been obscured for privacy reasons.

Note: Larger version of Figure 4 available here.

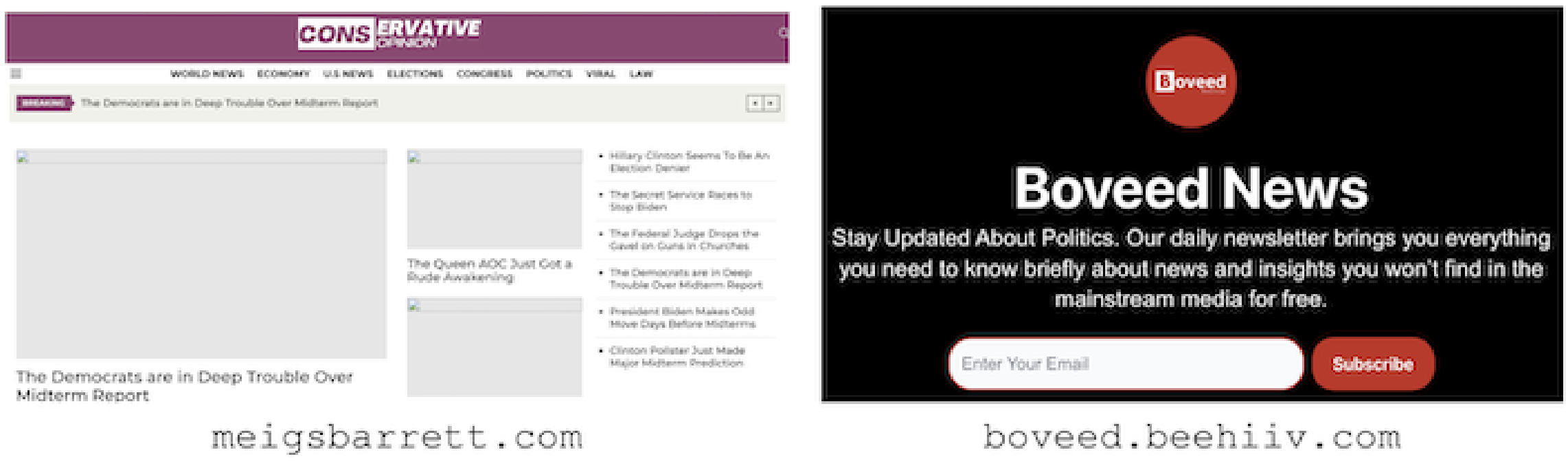

**Figure 5:** Media outlets that 15 users promote in their bio. We report a snapshot of meigsbarrett.com from the Internet Archive, as it is not reachable anymore. Interestingly, boveed.beehiiv.com links to an X account (@Boveedmedia) that is suspended at the time of writing (25 October 2024), and to the YouTube channel @MediaOpinion0, also promoted by other suspicious Web sites (*cf.*, Figure 6(b)) and several X accounts in their profile description (*cf.*, 1).

Note: Larger version of Figure 5 available here.

We tracked the actions of this coordinated network over time, examining their sharing activities from May to July. Through a semi-automated analysis of the content shared by these accounts, their profile, and the URLs they promote, we uncovered a cross-platform information operation involving coordinated accounts across X, YouTube, and the broader Web, promoting conservative content and the Republican presidential candidate (Figure 1). Many coordinated accounts displayed nearly identical profile pictures and biographies (Figure 4), which often link to the same external Web sites (Figure 5) — another signal of potential coordination. These accounts promoted a YouTube channel and six mock news outlets, four of which published identical content (Figure 6). Together, these findings offer compelling evidence of coordinated inauthentic behavior among the accounts detected by our model. It is worth noting that our model analyzes the shared URLs without making any assumptions or distinctions regarding their credibility or political orientation. The identification of coordinated accounts is entirely agnostic to the nature of the domains, focusing solely on similarities in user-sharing patterns. As a result, the Web domains promoted by coordinated accounts were identified after their detection, confirming the malicious and inauthentic nature of their sharing activity.

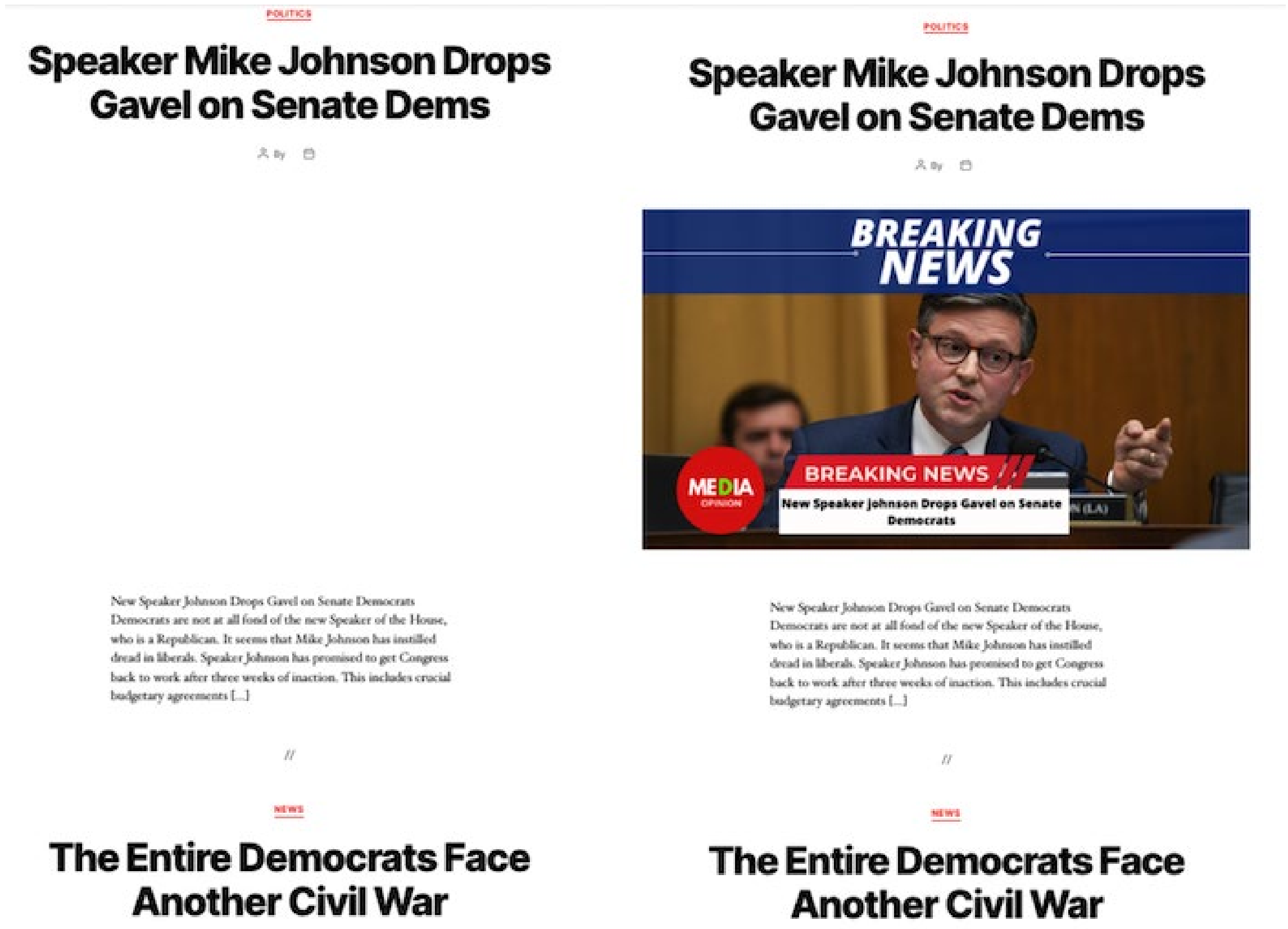

(a) patriotvoice.site (b) digmedia.life

**Figure 6:** Snapshots of two identical Web sites promoting biased narratives. Web Archive did not store the images for patriotvoice.site.

Note: Larger version of Figure 6 available here.

***Characterizing content promoted by inauthentic actors***

In this section, we describe our main findings by first scrutinizing the accounts identified as part of the IO, and then we shift our attention to the content promoted by the IO.

*Account-level analysis*. We examined the biographies and profile pictures of the 34 suspicious accounts, uncovering a high degree of coordination. Two accounts in the coordinated network shared the same profile and cover images (see Figure 1). Notably, 15 accounts shared the same phrase in their bio description: “Are you tired of fake news? clik on the link below”. Additionally, the links in their bios direct users to suspicious domains hosting alternative media outlets (meigsbarrett.com or boveed.beehiiv.com). Interestingly, the boveed.beehiiv.com Web site includes a link to the YouTube channel @MediaOpinion0 (also promoted by two X accounts in their profile), and to the X account @Boveedmedia (suspended at the time of writing, 25 October 2024). While meigsbarrett.com is currently down, boveed.beehiiv.com and the YouTube channel @MediaOpinion0 remain active.

Moreover, several accounts featured identical hashtags in the same exact sequence — (#MAGA2024 #Trump2024TheOnlyChoice #Trump2024) and (#TRUMP #MAGA, #TRUMPWON)— or similar opening statements like “Trump supporter voice” in their biography. Figure 4 illustrates three profile information of these coordinated accounts. We report the subset of suspended users in the Appendix. We detect a set of suspicious accounts sharing a conservative (meigsbarrett.com) and an alternative (boveed.beehiiv.com) media outlet in their bios. Two users also report the link to the YouTube account @MediaOpinion0 of Boveed News (boveed.beehiiv.com). We show the homepage of the YouTube account in Figure 7.

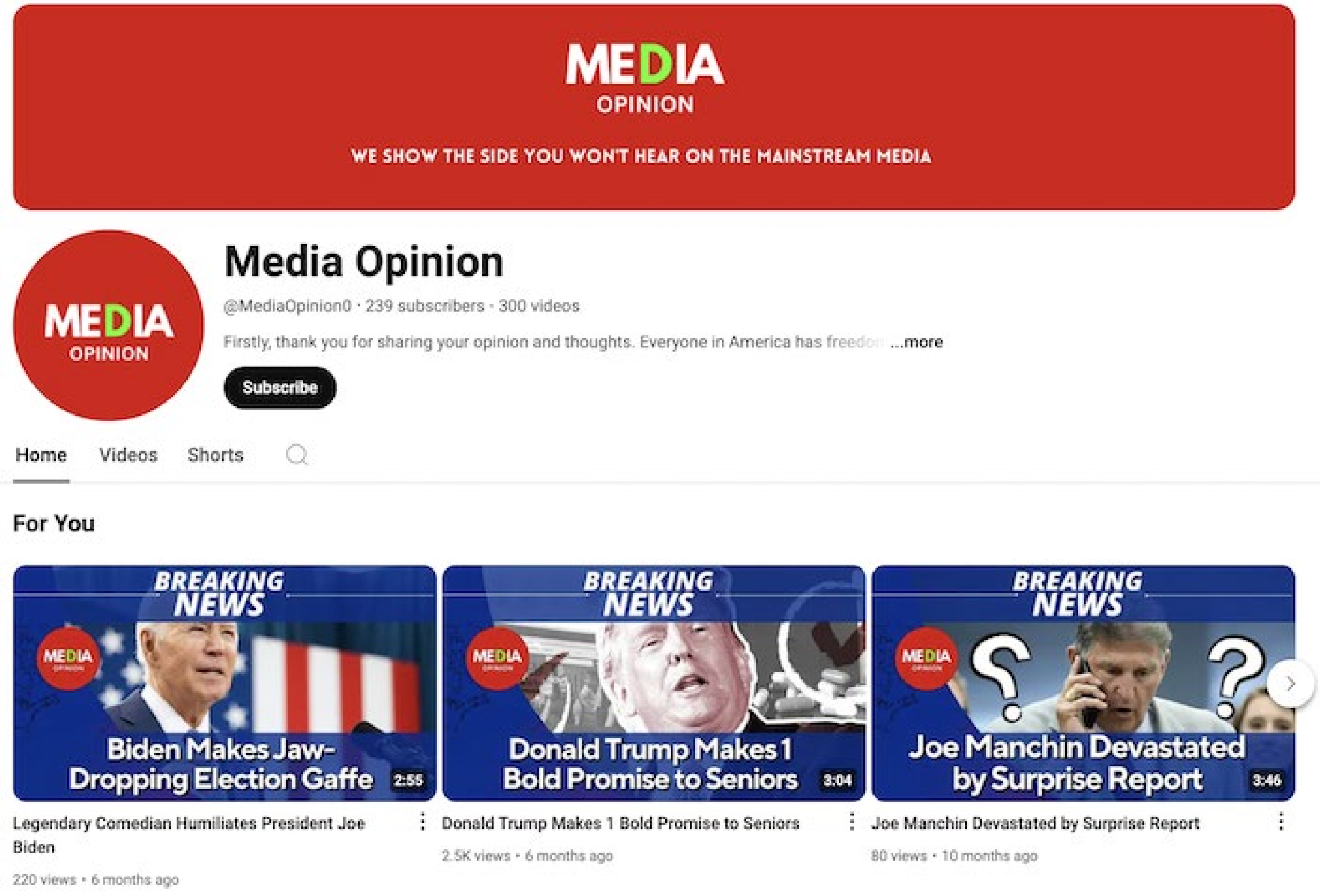

**Figure 7:** YouTube channel that was linked by two of the suspicious users in their bio. This same channel is referenced in the boveed.beehiiv.com as its official YouTube account.

Note: Larger version of Figure 7 available here.

*Content-level analysis*. In this section, we focus on the content that these coordinated actors shared on their feeds. This includes tweets pointing to external Web domains or other social media platforms, or containing AI images and AI-fabricated multimedia content.

*Web domains*. The accounts systematically shared links to five external Web sites, which are detailed in Table 2. These sites hosted articles promoting sensationalist, inflammatory, or biased narratives, with headlines such as “Since Biden Took Office Over 10 Million Illegal Crossed the Border”, “President Biden’s App Frees Thousands from Terrorism-Linked Nations into U.S.” or “Trump Vows to Reveal What Democrats Want Hidden Forever!” Notably, patriotvoice.site, conservativeon.site, americanvoice.site, and digmedia.life are clones, containing identical content (see Figure 6 and Appendix). Additionally, we used the API from apivoid.com to identify unsafe Web sites. This service detects malicious IP addresses and provides a trust score for each site under inspection. Trust scores below 50/100 indicate that a Web site may be unsafe or untrustworthy. For reference, Web sites like cnn.com and foxnews.com receive a score of 100/100. All sites promoted by coordinated users scored below 65, indicating low levels of trust and safety. Notably, all of these Web sites have since been taken down and are no longer accessible.

| Domain | Count | Trust Score | Trust Alert | Available Online |
|---|---|---|---|---|
| newsartificial.com | 9,236 | 65/100 | | ✗ |
| patriotvoice.site | 4,489 | 35/100 | ✓ | ✗ |
| conservativeon.site | 1,213 | 5/100 | ✓ | ✗ |
| americanvoice.site | 405 | 35/100 | ✓ | ✗ |
| digmedia.life | 125 | 60/100 | | ✗ |
| freemedia.live | 41 | 60/100 | | ✗ |

**Table 2:** Count statistics of the domains shared by coordinated actors. A trust alert is triggered for scores below 50/100, indicating that a Web site is unsafe or untrustworthy.

*AI-generated content*. We also identified the use of AI-generated content, with four accounts sharing AI-generated images. This assessment was conducted by two annotators who manually inspected the images shared by coordinated accounts. Their full agreement in identifying AI-generated content provides confidence in the reliability of this evaluation. We report three examples of AI-generated content in Figure 8. These images represent strong, emotionally charged depictions of Joe Biden and Donald Trump, with clear biases and intended messaging. Biden is portrayed mockingly, with insinuations of weakness or dependence on external support (as implied by the “pill” symbol). Trump, on the other hand, is depicted as powerful, victorious, and dominant, using both direct language (“9-0 Bitc**s”) and symbolic imagery (lion and military strength) to emphasize these traits.

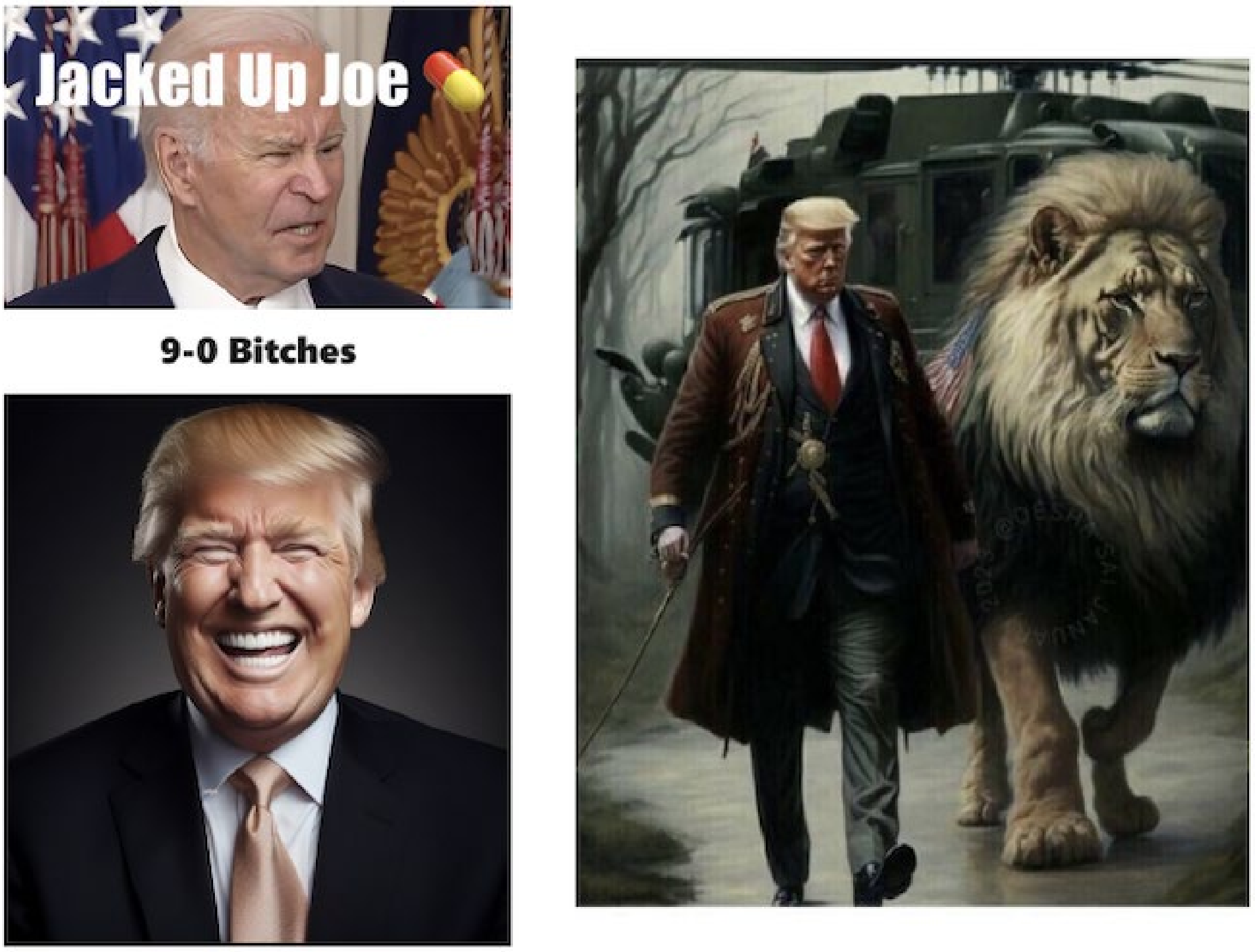

**Figure 8:** Examples of AI-generated content that was shared by users participating in the IO.

*Duplicated images*. Lastly, we observed three separate users posting the same image independently, rather than as part of a single cascade. This behavior, illustrated in Figure 9, reinforces the idea of strong coordination between these users. The image portrays a conservative narrative, showing an alleged support of the black community to the Republican presidential candidate. Interestingly, the duplicated image shown in Figure 9 traces back to a Reddit post from four years ago [2].

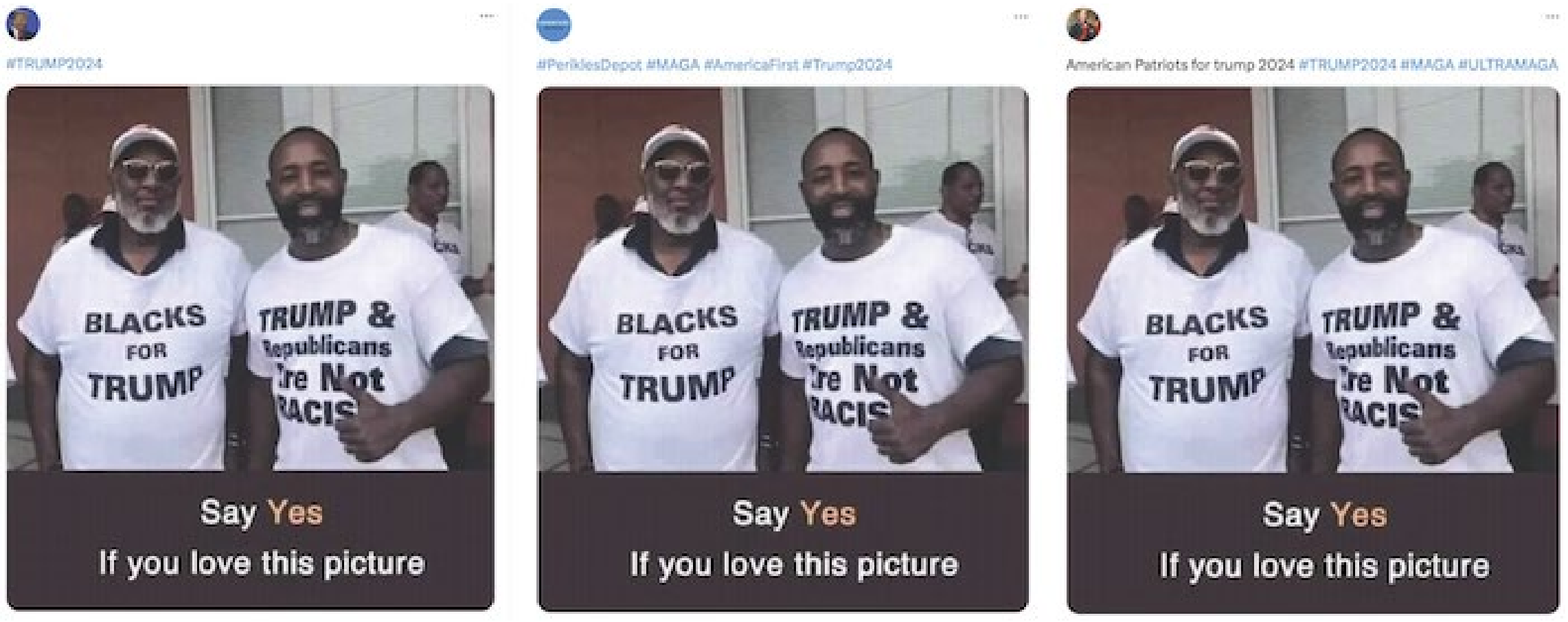

**Figure 9:** Three duplicated posts with the same media from three different IO actors' accounts.

Note: Larger version of Figure 9 available here.

*Cross-platform*. In addition to X/Twitter, coordinated users promoted several Web sites and the YouTube channel @MediaOpinion0, associated with the Web site Boveed News. We provide an extract of the YouTube channel homepage in Figure 10. Interestingly, the logo of this YouTube channel appears on the clone Web sites promoted by the coordinated network (see Figure 6), indicating multiple efforts to promote @MediaOpinion0 via several Web sites and X accounts. Although some of the Web sites have been taken down, @MediaOpinion0 is still active on YouTube, with over 30K engagements received so far. Overall, these findings surface the evidence of a cross-platform IO spanning X, YouTube, and the Web.

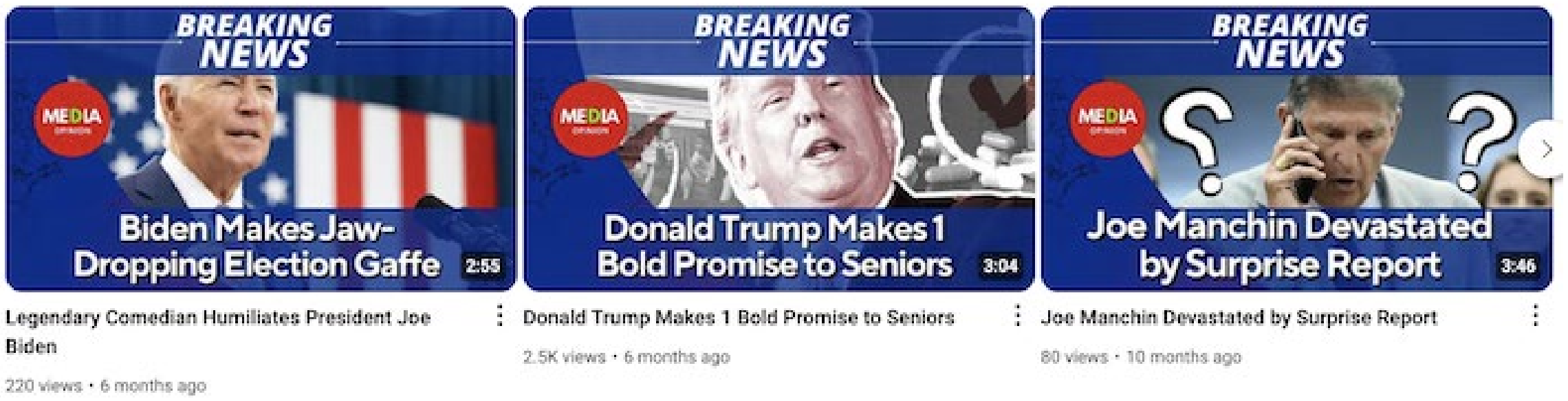

**Figure 10:** Extracts from the homepage of the YouTube channel linked in some IO actors' bios.

Note: Larger version of Figure 10 available here.

***Temporal analysis of inauthentic actors***

Following the identification and characterization of the coordinated accounts, we first detect other suspicious users in June and July who exhibit exceptionally high similarity to those in the coordinated network. Next, we analyze the temporal activity of all these accounts throughout the entire time frame. Finally, we assess whether these coordinated actions gained traction among the broader user base.

By analyzing user activity in June and July, we identify new accounts that exhibit extremely similar co-URL patterns (above the 90th percentile of the similarity distribution) to the coordinated accounts detected in May. This analysis reveals five additional highly similar accounts active in June and one in July, all of which were manually inspected as in previous assessments. This results in a total of 40 coordinated accounts operating from May to July. This subset remains relatively small due to the conservative nature of our approach, which is designed to minimize false positives, as previously mentioned.

We then analyze the temporal activity of coordinated accounts from May to July, focusing on the number of tweets that link to the suspicious domains listed in Table 2. Figure 11 (Top) displays the volume of tweets pointing to the top four suspicious Web sites, disentangled by Web domain. We do not consider digmedia.life and freemedia.live due to their limited promotion. The results show evident synchronization among accounts promoting patriotvoice.site, conservativeon.site, and americanvoice.site, with a noticeable increase in sharing activity after mid-June, followed by a sharp decline after mid-July, likely due to sites being taken down. It is important to note that this synchronization occurs across three Web sites, which are identical clones of each other (see Appendix). At the same time, the accounts increasingly promote content from newsartificial.com, with only a slight decrease in activity observed toward the end of June.

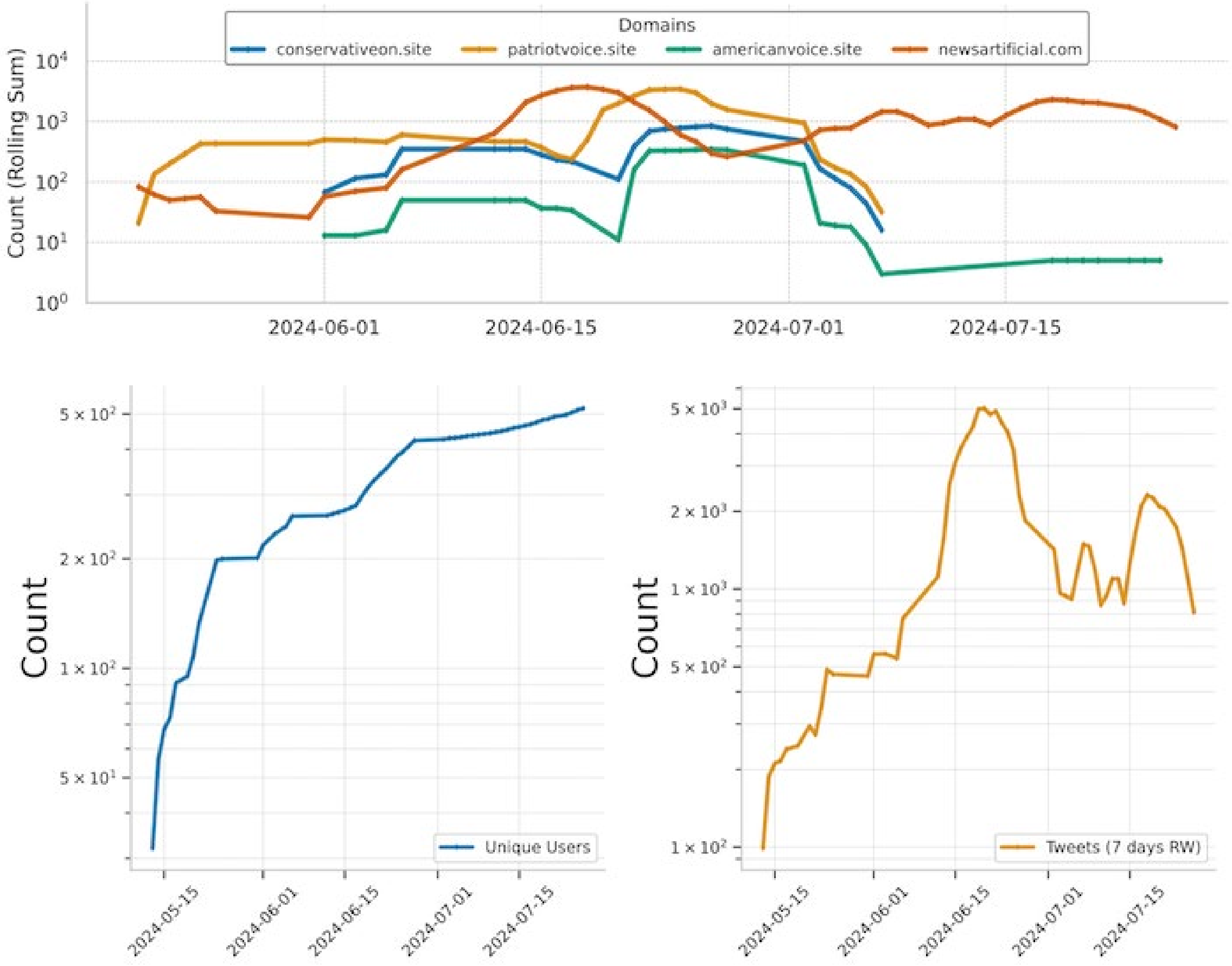

**Figure 11:** (Top) Volume of tweets shared by coordinated accounts across the top Web domains. (Bottom-left) Total engagement received by the top domains. (Bottom-right) Number of shares by the entire user base on the top domains. All time series are smoothed with a seven-day rolling window.

Note: Larger version of Figure 11 available here.

Finally, we examine the broader engagement these promoted domains have garnered among the wider user base. The bottom left panel of Figure 11 shows the cumulative number of unique users who shared at least one of the suspicious web domains over time. By 1 July, over 500 accounts had shared these domains — far exceeding the 40 accounts in the coordinated network. This rise is also evident in the volume of tweets embedding the suspicious Web sites, averaging 5,600 tweets per day. Although this activity declines in late June, it stabilizes at around 1,000 shares per day thereafter (Figure 11, bottom right), suggesting that the campaign gained significant momentum and maintained a consistent level of engagement despite the initial drop.

## Conclusions

In this study, we employed state-of-the-art scalable computational models to detect prominent IO activities across large social media platforms.

Looking at data from May to July 2024, we revealed a community of coordinated inauthentic accounts exhibiting IO-like activity. These accounts demonstrated striking similarities at the user level, including the adoption of identical or highly similar profile images and bios. They also shared overlapping external Web site domains and posted nearly identical content. Notably, some of these accounts featured AI-generated content in their original posts, but only 25 percent of these accounts have been suspended so far (25 October 2024).

The activities of this coordinated network extend beyond X and are closely linked to duplicated, suspicious Web sites, all of which promote a YouTube channel that is currently garnering tens of thousands of engagements. Together, this signals the emergence of a cross-platform effort to manipulate online audiences in the lead-up to the election, which gained considerable traction within the X ecosystem.

These findings underscore the critical role that scalable detection methods play in monitoring social media abuse. Their utility extends beyond individual platforms, impacting cross-platform (mis)information diffusion and the broader trustworthiness of the information ecosystem. The techniques we employed offer a powerful means to identify and mitigate the risks posed by IO campaigns, which continue to threaten the integrity of democratic discourse online, and by extension the upholding of fair political elections.

*Limitations*. While this study provides important evidence and characterizes cross-platform coordinated activity related to the 2024 U.S. presidential election, it is not without limitations. Our detection of malicious actors relies primarily on link-sharing patterns, which may overlook other forms of coordinated behavior that do not involve link dissemination. For example, certain actors may employ more subtle or varied coordination strategies, such as synchronized messaging or similar content sharing (Luceri, *et al.*, 2024; Pacheco, *et al.*, 2021). As a result, it is likely that our analysis underestimates the overall prevalence and complexity of coordinated activities originating on Twitter. Future work should explore additional dimensions of coordination to provide a more comprehensive understanding of these operations.

### About the team

The 2024 Election Integrity Initiative is led by Emilio Ferrara and Luca Luceri and carried out by a collective of USC students and volunteers whose contributions are instrumental to enable these studies. The authors are grateful to the following HUMANS Lab's members for their tireless efforts on this project: **Ph.D. students**: Charles 'Duke' Bickham, Leonardo Blas, Eun Cheol Choi, Priyanka Dey, Gabriela Pinto, Siyi Zhou. **Masters' students**: Ashwin Balasubramanian, Sneha Chawan, Vishal Reddy Chintham, Srilatha Dama, Yashvi Ashok Hiranandani, Joyston Menez, Jinhu Qi, Ameen Qureshi, Namratha Sairam, Tanishq Salkar, Srivarshan Selvaraj, Kashish Atit Shah, Gokulraj Varatharajan, Reuben Varghese. Undergraduate students: Isabel Epistelomogi, Collin Hargreaves, Saborni Kundu, Grace Li, Richard Peng, Brian Ramirez-Gonzalez, Christina You, Vito Zou. **Other collaborators**: Dr. Keith Burghardt.

### About the authors

**Marco Minici** is a Ph.D. student at the University of Pisa and ICAR-CNR (Institute for High Performance Computing and Networking).
E-mail: marco [dot] minici [at] icar [dot] cnr [dot] it

**Federico Cinus** is a Ph.D. student in data science at La Sapienza Università di Roma and CENTAI.
E-mail: cinus [at] diag [dot] uniroma1 [dot] it

**Luca Luceri** is research assistant professor of computer science and lead scientist at the Information Sciences Institute at the University of Southern California.
E-mail: lluceri [at] isi [dot] edu

**Emilio Ferrara** is a professor of computer science and communication at the University of Southern California.
E-mail: emiliofe [at] usc [dot] edu

### Authors' contributions

Marco Minici and Federico Cinus contributed equally to this work.

### Funding

This study has not been funded, sponsored, supported, or endorsed by any political party, funding agency, or any other organization.

### Notes

1. https://x.com/home

2. https://www.reddit.com/r/TheRightCantMeme/comments/h9ihh8/see_these_black_people_speak_for_all_black_people/

**References**

Isabelle Augenstein, Timothy Baldwin, Meeyoung Cha, Tanmoy Chakraborty, Giovanni Luca Ciampaglia, David Corney, Renee DiResta, Emilio Ferrara, Scott Hale, Alon Halevy, Heng Ji, Filippo Menczer, Ruben Miguez, Preslav Nakov, Dietram Scheufele, Shivam Sharma, and Giovanni Zagni, 2024. "Factuality challenges in the era of large language models and opportunities for fact-checking," *Nature Machine Intelligence*, volume 6, pp. 852–863.
doi: https://doi.org/10.1038/s42256-024-00881-z, accessed 29 October 2024.

Adam Badawy, Aseel Addawood, Kristina Lerman, and Emilio Ferrara, 2019. "Characterizing the 2016 Russian IRA influence campaign," *Social Network Analysis and Mining*, volume 9, article number 31.
doi: https://doi.org/10.1007/s13278-019-0578-6, accessed 29 October 2024.

Christopher A. Bail, Brian Guay, Emily Maloney, Aidan Combs, D. Sunshine Hillygus, Friedolin Merhout, Deen Freelon, and Alexander Volfovsky, 2020. "Assessing the Russian Internet Research Agency's impact on the political attitudes and behaviors of American Twitter users in late 2017," *Proceedings of the National Academy of Sciences*, volume 117, number 1, pp. 243–250.
doi: https://doi.org/10.1073/pnas.1906420116, accessed 29 October 2024.

Alexandre Bovet and Hernán A Makse, 2019. "Influence of fake news in Twitter during the 2016 US presidential election," *Nature Communications*, volume 10, article number 7.
doi: https://doi.org/10.1038/s41467-018-07761-2, accessed 29 October 2024.

Emily Chen and Emilio Ferrara, 2023. "Tweets in time of conflict: A public dataset tracking the Twitter discourse on the war between Ukraine and Russia," *Proceedings of the Seventeenth International AAAI Conference on Web and Social Media*, volume 17, pp. 1,006–1,013.
doi: https://doi.org/10.1609/icwsm.v17i1.22208, accessed 29 October 2024.

Ashok Deb, Luca Luceri, Adam Badawy, and Emilio Ferrara, 2019, "Perils and challenges of social media and election manipulation analysis: The 2018 US midterms," *WWW '19: Companion Proceedings of The 2019 World Wide Web Conference*, pp. 237–247.
doi: https://doi.org/10.1145/3308560.3316486, accessed 29 October 2024.

Priyanka Dey, Luca Luceri, and Emilio Ferrara, 2024. "Coordinated activity modulates the behavior and emotions of organic users: A case study on tweets about the Gaza conflict," *WWW '24: Companion Proceedings of the ACM Web Conference 2024*, pp. 682–685.
doi: https://doi.org/10.1145/3589335.3651483, accessed 29 October 2024.

Emilio Ferrara, 2024a. "Charting the landscape of nefarious uses of generative artificial intelligence for online election interference," *arXiv*:2406.01862 (4 June).
doi: https://doi.org/10.48550/arXiv.2406.01862, accessed 29 October 2024.

Emilio Ferrara, 2024b. "GenAI against humanity: Nefarious applications of generative artificial intelligence and large language models," *Journal of Computational Social Science*, volume 7, pp. 549–569.
doi: https://doi.org/10.1007/s42001-024-00250-1, accessed 29 October 2024.

Emilio Ferrara, 2020. "Bots, elections, and social media: A brief overview," In: Kai Shu, Suhang Wang, Dongwon Lee, and Huan Liu (editors). *Disinformation, misinformation, and fake news in social media: Emerging research challenges and opportunities*. Cham, Switzerland: Springer, pp. 95–114.
doi: https://doi.org/10.1007/978-3-030-42699-6_6, accessed 29 October 2024.

Emilio Ferrara, Herbert Chang, Emily Chen, Goran Muric, and Jaimin Patel, 2020. "Characterizing social media manipulation in the 2020 U.S. presidential election," *First Monday*, volume 25, number 11.
doi: https://doi.org/10.5210/fm.v25i11.11431, accessed 29 October 2024.

Nicholas A. Gabriel, David A. Broniatowski, and Neil F. Johnson, 2023. "Inductive detection of influence operations via graph learning," *Scientific Reports*, volume 13, article number 22571.
doi: https://doi.org/10.1038/s41598-023-49676-z, accessed 29 October 2024.

Fabio Giglietto, Nicola Righetti, Luca Rossi, and Giada Marino, 2020. "It takes a village to manipulate the media: Coordinated link sharing behavior during 2018 and 2019 Italian elections," *Information, Communication & Society*, volume 23, number 6, pp. 867–891.
doi: https://doi.org/10.1080/1369118X.2020.1739732, accessed 29 October 2024.

Xiaobo Guo and Soroush Vosoughi, 2022. "A large-scale longitudinal multimodal dataset of state-backed information operations on Twitter," *Proceedings of the Sixteenth International AAAI Conference on Web and Social Media*, volume 16, pp. 1,245–1,250.
doi: https://doi.org/10.1609/icwsm.v16i1.19375, accessed 29 October 2024.

Kristina Hristakieva, Stefano Cresci, Giovanni Da San Martino, Mauro Conti, and Preslav Nakov, 2022. "The spread of propaganda by coordinated communities on social media," *WebSci '22: Proceedings of the 14th ACM Web Science Conference 2022*, pp. 191–201. doi: https://doi.org/10.1145/3501247.3531543, accessed 29 October 2024.

Luca Luceri, Eric Boniardi, and Emilio Ferrara, 2024. "Leveraging large language models to detect influence campaigns on social media," *WWW '24: Companion Proceedings of the ACM Web Conference 2024*, pp. 1,459–1,467. doi: https://doi.org/10.1145/3589335.3651912, accessed 29 October 2024.

Luca Luceri, Felipe Cardoso, and Silvia Giordano, 2021. "Down the bot hole: Actionable insights from a one-year analysis of bot activity on Twitter," *First Monday*, volume 26, number 3. doi: https://doi.org/10.5210/fm.v26i3.11441, accessed 29 October 2024.

Luca Luceri, Ashok Deb, Adam Badawy, and Emilio Ferrara, 2019. "Red bots do it better: Comparative analysis of social bot partisan behavior," *WWW '19: Companion Proceedings of The 2019 World Wide Web Conference*, pp. 1,007–1,012. doi: https://doi.org/10.1145/3308560.3316735, accessed 29 October 2024.

Luca Luceri, Ashok Deb, Silvia Giordano, and Emilio Ferrara, 2019. "Evolution of bot and human behavior during elections," *First Monday*, volume 24, number 9. doi: https://doi.org/10.5210/fm.v24i9.10213, accessed 29 October 2024.

Luca Luceri, Valeria Pantè, Keith Burghardt, and Emilio Ferrara, 2024. "Unmasking the web of deceit: Uncovering coordinated activity to expose information operations on Twitter," *WWW '24: Proceedings of the ACM Web Conference 2024*, pp. 2,530–2,541. doi: https://doi.org/10.1145/3589334.3645529, accessed 29 October 2024.

Thomas Magelinski, Lynnette Ng, and Kathleen Carley, 2022. "A synchronized action framework for detection of coordination on social media," *Journal of Online Trust & Safety*, volume 1, number 2. doi: https://doi.org/10.54501/jots.v1i2.30, accessed 29 October 2024.

Diego A. Martin, Jacob N. Shapiro, and Michelle Nedashkovskaya, 2019. "Recent trends in online foreign influence efforts," *Journal of Information Warfare*, volume 18, number 3, pp. 15–48.

Meta, 2024. "Integrity reports, second quarter 2024" (3 October), at https://transparency.meta.com/en-us/integrity-reports-q2-2024/, accessed 29 October 2024.

Microsoft Threat Intelligence, 2024. "Iran surges cyber-enabled influence operations in support of Hamas," at https://www.microsoft.com/en-us/security/security-insider/intelligence-reports/iran-surges-cyber-enabled-influence-operations-in-support-of-hamas, accessed 29 October 2024.

Maximilian Mozes, Xuanli He, Bennett Kleinberg, and Lewis D. Griffin, 2023. "Use of LLMs for illicit purposes: Threats, prevention measures, and vulnerabilities," *arXiv*:2308.12833 (24 August). doi: https://doi.org/10.48550/arXiv.2308.12833, accessed 29 October 2024.

Leonardo Nizzoli, Serena Tardelli, Marco Avvenuti, Stefano Cresci, and Maurizio Tesconi, 2021. "Coordinated behavior on social media in 2019 UK general election," *Proceedings of the International AAAI Conference on Web and Social Media*, volume 15. pp. 443–454. doi: https://doi.org/10.1609/icwsm.v15i1.18074, accessed 29 October 2024.

OpenAI, 2024. "AI and covert influence operations: Latest trends," at https://downloads.ctfassets.net/kftzwdyauwt9/5IMxzTmUclSOAcWUXbkVrK/3cfab518e6b10789ab8843bcca18b633/Threat_Intel_Report.pdf, accessed 29 October 2024.

Diogo Pacheco, Alessandro Flammini, and Filippo Menczer, 2020. "Unveiling coordinated groups behind white helmets disinformation," *WWW '20: Companion Proceedings of the Web Conference 2020*, pp. 611–616. doi: https://doi.org/10.1145/3366424.3385775, accessed 29 October 2024.

Diogo Pacheco, Pik-Mai Hui, Christopher Torres-Lugo, Bao Tran Truong, Alessandro Flammini, and Filippo Menczer, 2021. "Uncovering coordinated networks on social media: Methods and case studies," *Proceedings of the International AAAI Conference on Web and Social Media*, volume 15, pp. 455–466. doi: https://doi.org/10.1609/icwsm.v15i1.18075, accessed 29 October 2024.

Francesco Pierri, Luca Luceri, Nikhil Jindal, and Emilio Ferrara, 2023. "Propaganda and misinformation on Facebook and Twitter during the Russian invasion of Ukraine," *WebSci '23: Proceedings of the 15th ACM Web Science Conference 2023*, pp. 65–74. doi: https://doi.org/10.1145/3578503.3583597, accessed 29 October 2024.

Iacopo Pozzana and Emilio Ferrara, 2020. "Measuring bot and human behavioral dynamics," *Frontiers in Physics*, volume 8, 125. doi: https://doi.org/10.3389/fphy.2020.00125, accessed 29 October 2024.

Karishma Sharma, Emilio Ferrara, and Yan Liu, 2022. "Characterizing online engagement with disinformation and conspiracies in the 2020 US presidential election," *Proceedings of the International AAAI Conference on Web and Social Media*, volume 16, pp. 908–919.
doi: https://doi.org/10.1609/icwsm.v16i1.19345, accessed 29 October 2024.

Karishma Sharma, Yizhou Zhang, Emilio Ferrara, and Yan Liu, 2021. "Identifying coordinated accounts on social media through hidden influence and group behaviours," *KDD '21: Proceedings of the 27th ACM SIGKDD Conference on Knowledge Discovery & Data Mining*, pp. 1,441–1,451.
doi: https://doi.org/10.1609/icwsm.v16i1.19345, accessed 29 October 2024.

Felipe Bonow Soares, Anatoliy Gruzd, and Philip Mai, 2023. "Falling for Russian propaganda: Understanding the factors that contribute to belief in pro-Kremlin disinformation on social media," *Social Media + Society* (22 December).
doi: https://doi.org/10.1177/20563051231220330, accessed 29 October 2024.

Kate Starbird, Ahmer Arif, and TomWilson, 2019. "Disinformation as collaborative work: Surfacing the participatory nature of strategic information operations," *Proceedings of the ACM on Human-Computer Interaction*, volume 3, number CSCW, article number 127, pp. 1–26.
doi: https://doi.org/10.1145/3359229, accessed 29 October 2024.

Massimo Stella, Emilio Ferrara, and Manlio De Domenico, 2018. "Bots increase exposure to negative and inflammatory content in online social systems," *Proceedings of the National Academy of Sciences*, volume 115, number 49 (20 November), pp. 12,435–12,440.
doi: https://doi.org/10.1073/pnas.1803470115, accessed 29 October 2024.

Serena Tardelli, Leonardo Nizzoli, Maurizio Tesconi, Mauro Conti, Preslav Nakov, Giovanni Da San Martino, and Stefano Cresci, 2024. "Temporal dynamics of coordinated online behavior: Stability, archetypes, and influence," *Proceedings of the National Academy of Sciences*, volume 121, number 20 (6 May), e2307038121.
doi: https://doi.org/10.1073/pnas.2307038121, accessed 29 October 2024.

U.S. Department of State. Office of the Spokesperson, 2024. "Alerting the world to RT's global covert activities" (13 September), at https://www.state.gov/alerting-the-world-to-rts-global-covert-activities/, accessed 29 October 2024.

Luis Vargas, Patrick Emami, and Patrick Traynor, 2020. "On the detection of disinformation campaign activity with network analysis," *CCSW'20: Proceedings of the 2020 ACM SIGSAC Conference on Cloud Computing Security Workshop*, pp. 133–146.
doi: https://doi.org/10.1145/3411495.3421363, accessed 29 October 2024.

Padinjaredath Suresh Vishnuprasad, Gianluca Nogara, Felipe Cardoso, Stefano Cresci, Silvia Giordano, and Luca Luceri, 2024. "Tracking fringe and coordinated activity on Twitter leading up to the US Capitol attack," *Proceedings of the International AAAI Conference on Web and Social Media,*, volume 18, pp. 1,557–1,570.
doi: https://doi.org/10.1609/icwsm.v18i1.31409, accessed 29 October 2024.

Derek Weber and Frank Neumann, 2021. "Amplifying influence through coordinated behaviour in social networks," *Social Network Analysis and Mining*, volume 11, article number 11.
doi: https://doi.org/10.1007/s13278-021-00815-2, accessed 29 October 2024.

Asta Zelenkauskaite, Pihla Toivanen, Jukka Huhtamäki, and Katja Valaskivi, 2021. "Shades of hatred online: 4chan duplicate circulation surge during hybrid media events," *First Monday*, volume 26, number 1.
doi: https://doi.org/10.5210/fm.v26i1.11075, accessed 29 October 2024.

## Appendix

*Keywords*. Joe Biden, Donald Trump, 2024 US Elections, US Elections, 2024 Elections, 2024 Presidential Elections, Biden, Joe Biden, Joseph Biden, Biden2024, Donald Trump, Trump2024, trumpsupporters, trumptrain, republicansoftiktok, conservative, MAGA, KAG, GOP, CPAC, Nikki Haley, Ron DeSantis , RNC, democratsoftiktok, thedemocrats, DNC, Kamala Harris, Marianne Williamson, Dean Phillips, williamson2024, phillips2024, Democratic party, Republican party, Third Party, Green Party, Independent Party, No Labels, RFK Jr, Roberty F. Kennedy Jr. , Jill Stein, Cornel West, ultramaga, voteblue2024, letsgobrandon, bidenharris2024, makeamericagreatagain, Vivek Ramaswamy, JD Vance, Assassination, Tim Walz, WWG1WGA.

*Social media domains*. We separated popular social media domains from general domains, presenting the latter in the main text. Here, we report the top five social media-related domains included in the dataset.

*News articles*. The users shared many links to four conservative media outlets, listed in Table 2. We report in Figure 12 three extracts of news articles shared on these Web sites.

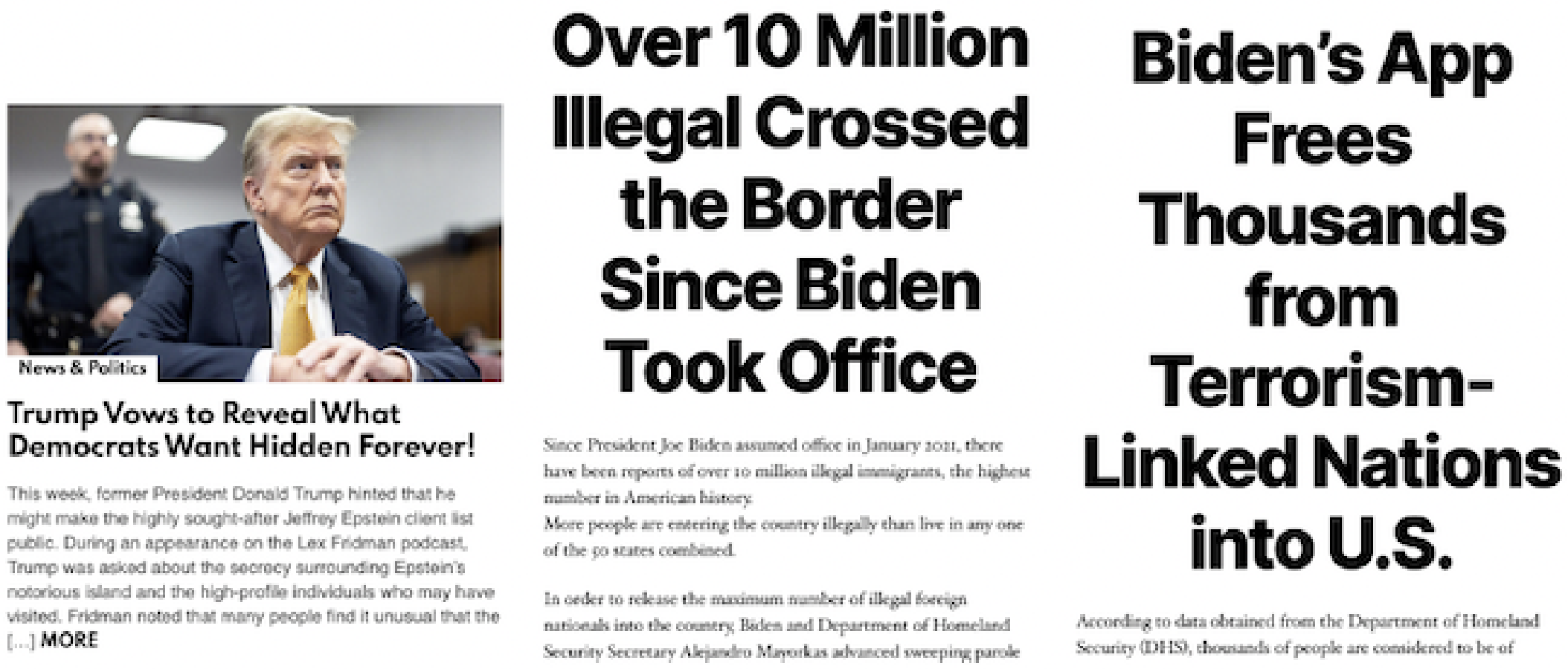

**Figure 12:** Extract of news articles present on the media outlets shared by highly coordinated users. The example on the left comes from newsartificial.com, the two on the right from patriotvoice.site.

*Suspended accounts*. We report the subset of highly coordinated accounts that are suspended at the time of writing (25 October 2024): 1768199419495690240, 1545059729902145538, 1765992285601480704, 1601247324231049218, 1574118779738476544, 1545800072113360896, 1764912986743885824, 1776131571080114177, 1765263292786700288, 1574096740696920065.

*Restricted accounts*. We report the presence of warnings for some coordinated accounts on X due to unusual activity (see the screenshot in Figure 13.

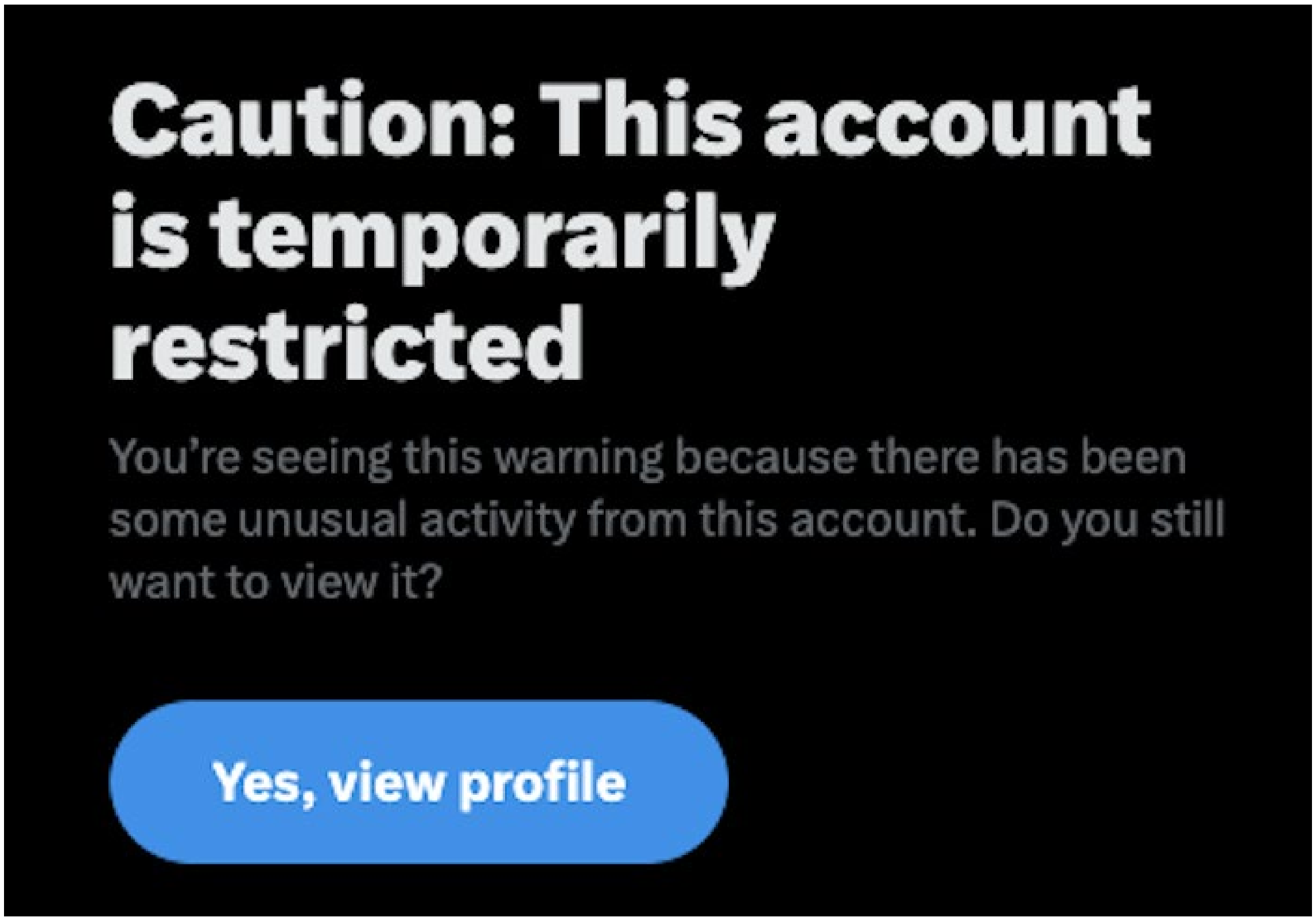

**Figure 13:** Warning message for restricted users due to unusual activity on X.

*Clone mock Web sites*. In the section "Characterizing content promoted by inauthentic actors," we present the domains promoted by the detected coordinated accounts. We also provide screenshots of four Web sites that share identical content and layout in Figure 14. Since these sites were taken down, we obtained the screenshots from the Web Archive. The links to these archived snapshots are as follows: patriotvoice.site, conservativeon.site, americanvoice.site, and digmedia.life.

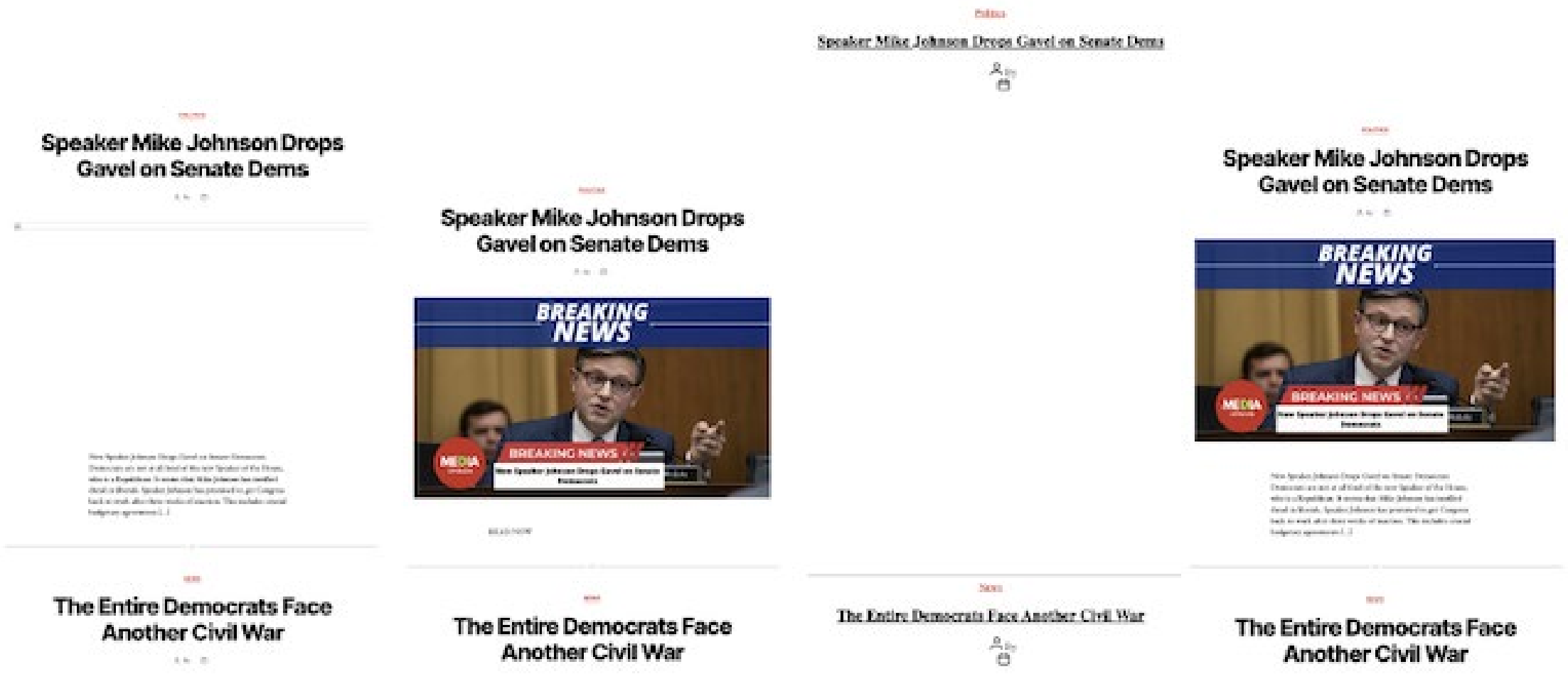

**Figure 14:** Snapshots of the front page of patriotvoice.site, conservativeon.site, americanvoice.site and digmedia.life. Web Archive stored the images only for conservativeon.site and digmedia.life.

---

**Editorial history**

---